\input harvmac

\def\a{\alpha}
\def\ah{{\hat\a}}

\def\b{\beta}

\def\bh{\hat\b}

\def\d{\delta}

\def\e{\epsilon}

\def\g{\gamma}

\def\l{\lambda}

\def\lb{{\overline\lambda}}
\def\o{\omega}

\def\ob{\overline\omega}
\def\O{\Omega}

\def\t{\theta}

\def\G{\Gamma}

\def\L{\Lambda}

\def\N{\nabla}
\def\Nb{\overline\nabla}

\def\p{\partial}
\def\pb{\overline\partial}
\def\Jb{\overline J}

\Title{ \vbox{\baselineskip12pt }}
{\vbox{\centerline{GGI Lectures on the}
\bigskip
\centerline{ Pure Spinor Formalism of the Superstring
 }}}

\bigskip
\centerline{Oscar A. Bedoya$^a$\foot{e-mail: abedoya@fma.if.usp.br}
and Nathan Berkovits$^b$\foot{email: nberkovi@ift.unesp.br}}
\bigskip
\centerline{\it $^a$Instituto de F\'{\i}sica,
Universidade de S\~ao Paulo} \centerline{\it 05315-970, S\~ao Paulo, SP, 
Brasil }
\centerline{\it $^b$Instituto de F\'{\i}sica Te\'orica,
UNESP-Universidade Estadual de S\~ao Paulo} 
\centerline{\it 01140-070, S\~ao Paulo, SP, Brasil }

\bigskip

\centerline{ Notes taken by Oscar A. Bedoya of lectures of Nathan Berkovits}
\centerline{in June 2009 at the Galileo Galilei Institute School}
\centerline{ ``New Perspectives in String Theory''}

\bigskip
\bigskip

\centerline{\bf Outline}

1. Introduction

2. $d=10$ Super Yang-Mills and Superparticle

3. Pure Spinor Superstring and Tree Amplitudes

4. Loop Amplitudes

5. Curved Backgrounds

6. Open Problems

\Date{}

\newsec{Introduction}

\subsec{Ramond-Neveu-Schwarz formalism}
The superstring in the RNS formalism has four different sectors. In
the NS GSO($+$) sector, there are the massless vector and massive states while
in the NS GSO($-$) there are the tachyon and massive modes. On the other
hand, in the R GSO($+$) sector, there are massless Weyl and massive
states, while in the R GSO($-$) there are anti-Weyl massless and massive
states. Although the GSO projection projects out the GSO($-$) part of
the spectrum, some processes (such as tachyon condensation) involve this
sector. 

The RNS formalism in the NS GSO($+$) and NS GSO($-$) sectors is
supersymmetric at the worldsheet level. For the
open string, it can be described by a superfield in two dimensions
\eqn\RNSX{{\bf X}^m (z,\kappa) = X^m (z) + \kappa \psi^m (z).}
In this formalism, can write vertex operators for the massless field
in the GSO($+$) sector
\eqn\vertexRNS{V = \int dz d\kappa (D {\bf X}^m )A_m ({\bf X}),}
where the derivative is $D = {\p \over{\p\kappa}} + \kappa {\p \over{\p
z}}$. 
The tachyon in the GSO($-$) sector can be described by the vertex
operator
\eqn\vertextachyon{V_T = \int dz d\kappa T({\bf X}) = \int dz (\psi \cdot
{\p \over{\p X}})T.}
For the R sector, a vertex operator can be written, but it is more
complicated, is not manifestly worldsheet supersymmetric,
and involves the spin field
\ref\FMS{D. Friedan, E. Martinec and S. Shenker, {\it
Conformal Invariance, Supersymmetry and String Theory, } Nucl Phys.
B271:93 1986.}
\eqn\vertexfermion{\Sigma^\a = e^{{1\over 2} \int \psi \psi}
e^{{1\over 2} \int \b\g}.}
Because of the complicated nature of the Ramond vertex operator, 
scattering amplitudes using the RNS
formalism have been computed up to
6 fermions at tree level \ref\kost{
V.A. Kostelecky, O. Lechtenfeld, S. Samuel, D. Verstegen, S. Watamura and
D. Sahdev, {\it The
Six-Fermion Amplitude in the Superstring,}
Phys. Lett. B183 (1987) 299.}, up to 4 fermions at one loop \ref\atick{
J. Atick and A. Sen, {\it
Covariant One-Loop Fermion Emission Amplitudes in Closed String Theories,}
Nucl. Phys. B293 (1987) 317.}
and, for 2-loops, the only RNS
computations involve
4 bosons and no fermions \ref\dhoker{E. D'Hoker and D. Phong,
{\it
Two-Loop Superstrings VI: Non-Renormalization Theorems and the 4-point 
Function,}
Nucl. Phys. B715 (2005) 3, 
hep-th/0501197.}.

For curved backgrounds, in the bosonic string case, the action can be
written as
\eqn\bosoniccurved{S = \int d^2 z g_{mn} \p X^m \pb X^n}
or with an antisymmetric field coupling $b_{mn}(X)$
\eqn\bosonicsigma{S = \int d^2 z (g_{mn} + b_{mn})\p X^m \pb X^n.}

There is an obvious generalization for the RNS formalism 
\eqn\RNScurved{S = \int d^2 z d^2 \kappa [g_{mn}({\bf X}) + b_{mn}({\bf
X})]D {\bf X}^m \bar D {\bf X}^n}
where $\bar D = {\p \over{\p\bar\kappa}} + \bar \kappa {\p \over{\p
\bar z}}$. This action for the NS-NS sector
can be obtained at the linearized level
as the product of two massless vector
states. But if one tries to describe the R-R sector
by naively introducing a term $\Sigma ^\a
\bar \Sigma^\b F_{\a\b} (X)$ to the action, where $\Sigma^\a$ is
the fermionic vertex operator introduced above, this term would
require picture changing operators since the back-reaction of the
R-R term would
not be in the same picture as the NS-NS term. Since picture-changing
is related to worldsheet superconformal invariance and
is only understood in on-shell NS-NS backgrounds, it is unclear
how to describe the RNS formalism in an R-R background.

If one computes amplitudes in the RNS formalism where all external states are in
the NS sector, there could be internal R states in the
loops. This means one has to sum over spin structures, which
complicates the computation of loop amplitudes. However, if one
computes amplitudes where all external states are in the GSO($+$) sector,
all internal states in the loops will also be GSO($+$). This suggests
one should try to describe the
superstring in a space-time supersymmetric way in which one only has
the GSO($+$) sector. 

The natural variables for the GSO($+$) sector are $X^m (z)$ for $m=
0,{\ldots} 9$ and $\t^\a (z)$ for $\a = 1,{\ldots} 16$, and the vertex
operators will be functions of $X^m$ and $\t^\a$. Space-time
supersymmetry transforms
\eqn\susy{\d \t^\a = \e^\a ,\,\,\,\, \d X^m = (\e \g^m \t).}
It will be important to fix the notation used. $\g^m _{\a\b}$ and
$(\g^m)^{\a\b}$ denotes
$16\times 16$ symmetric matrices which are the off-diagonal components
of the $32\times 32$ $\G^m$ matrices. Thus, the $\g^m$ matrices are
the analog of the Pauli matrices in 10 dimensions. They satisfy the
algebra $\g^{(m}_{\a\b} \g^{n)\b\g} = 2\eta^{mn}\d_\a {}^\g$. By
antisymmetrizing the product of 3 gamma matrices, one can check that
$(\g^{mnp})_{\a\b} = -(\g^{mnp})_{\b\a}$, while by antisymmetrizing
the product of 5 gamma matrices, one can check that
$(\g^{mnpqr})_{\a\b} = (\g^{mnpqr})_{\b\a}$. 

\subsec{Green-Schwarz formalism}

There is a classical description for the superstring using these variables known
as the Green-Schwarz formalism \ref\GSformalism{M. Green and J.
Schwarz, {\it Covariant Description of Superstrings,} Phys. Lett. B136
(1984) 367.}. In order to compute the spectrum one must impose 
the light-cone gauge.  
On the other hand, the light-cone gauge choice makes difficult
scattering amplitude computations, since some unphysical
singularities appear in
the worldsheet diagrams. Because of the hidden Lorentz
invariance, these unphysical singularities must cancel, however, this
is difficult to show explicitly. In any case, up to now only
4-point tree and one loop amplitudes have been 
explicitly computed using this formalism \ref\gst{M.B. Green
and J.H. Schwarz, {\it Supersymmetrical String Theories,}
Phys. Lett. B109 (1982) 444.}.

\subsec{Pure spinor formalism}

In these lectures, a new formalism for the superstring \ref\pureone
{N. Berkovits, {\it Super-Poincar\'e Covariant Quantization of the
Superstring,} JHEP 0004 (2000) 018, hep-th/0001035.} will be presented
which has made
progress on both computing scattering amplitudes and describing backgrounds in
a manifestly spacetime-supersymmetric manner.

1. Scattering amplitude computations:

It has been computed $N$-point tree
amplitudes with an arbitrary number of fermions \ref\Npttree{N. Berkovits
and B. Vallilo, {\it Consistency of super-Poincare covariant
superstring tree amplitudes,} JHEP 0007:015, 2000, [arXiv] hep-th/0004171.}, 
5-point one-loop
amplitudes with up to four fermions \ref\MafraStahn{C. Mafra and
C. Stahn, {\it The One-loop Open Superstring Massless Five-pint
Amplitude with the Non-Minimal Pure Spinor Formalism,} JHEP 0903:126,
2009, arXiv:0902.1539 [hep-th]. }, and 4-point two-loop amplitudes with
up to four fermions
\ref\twoloop{N. Berkovits and C. Mafra, {\it Equivalence of two-loop
superstring amplitudes in the pure spinor and RNS formalism,} Phys.
Revl. Lett. 96: 011602, 2006, hep-th/0509234.}\ref\BM{N. Berkovits and C.
Mafra, {\it Some Superstring Amplitude Computations with the
Non-Minimal Pure Spinor Formalism,} JHEP 0611 (2006) 079,
hep-th/0607187.}.

Beyond $2$-loops there are vanishing (non-renormalization) theorems
stating that beyond a certain loop order,
the effective action will not get contributions
containing a certain number of
derivatives of $R^4$ \ref\nrtheorems{N. Berkovits, {\it New higher-derivative
$R^4$ theorems,} Phys. Rev. Lett. 98:211601, 2007,
e-Print:hep-th/0609006.}. The proof relies on the counting of fermionic zero
modes which are related to space-time supersymmetry.
For $g\leq 6$, $\p^{2g} R^4$ is the lowest order term which appears
at genus $g$. If this statement could be extended for all $g$,
it would imply that $N=8$ $d=4$
supergravity is finite \ref\bern{
Z. Bern, J.J. Carrasco, L.J. Dixon, H. Johansson and R. Roiban, 
{\it The Ultraviolet Behavior of N=8 Supergravity at Four Loops,}
Phys. Rev. Lett. 103 (2009) 081301, arXiv:0905.2326 [hep-th].}
\ref\GRV{M. Green, J. Russo and P.
Vanhove, {\it Ultraviolet properties of maximal supergravity,} Phys.
Rev. Lett. 98:131602, 2007. e-Print: hep-th/0611273.}. However, it naively
appears
that $\p^{12} R^4$ terms are present for all $g\geq 6$, which implies
by dimensional arguments that $N=8$ $d=4$ sugra 
is divergent starting at 9 loops \GRV.

2. Ramond-Ramond backgrounds: 

In the pure spinor formalism, these backgrounds are no more complicated 
than NS-NS backgrounds. They are necessary to study the string in
$AdS_5 \times S^5$. Some work has been done in the GS formalism and 
$PSU(2,2|4)$ invariance in $AdS_5\times S^5$
plays the same role as super-Poincare
invariance in a flat background. So quantization in the GS formalism
requires breaking the manifest $PSU(2,2|4)$ invariance whereas quantization
in the pure spinor formalism preserves this symmetry.

Using the pure spinor formalism it has been shown that
strings in the $AdS_5\times S^5$
background are consistent at the quantum level to all
orders in $\a'$ \ref\Quantumconsistency{N. Berkovits, {\it Quantum
consistency of the superstring in $AdS_5\times S^5$ background,}
JHEP 0503:041, 2005. e-Print: hep-th/0411170.}. Non-local conserved
currents were constructed
\ref\BRP{I. Bena, J. Polchinski, R. Roiban, {\it Hidden
symmetries of the $AdS(5)\times S^5 $ superstring,} Phys. Rev. D69:
046002, 2004, arXiv[hep-th/0305116]}  
\ref\flatcurrents{B. Vallilo, {\it Flat currents in the classical
$AdS_5\times S^5$ pure spinor superstring,} JHEP 0403:037, 2004,
hep-th/0307018.}\ref\mik{
A. Mikhailov and S. Schafer-Nameki, 
{\it Algebra of transfer-matrices and 
Yang-Baxter equations on the string worldsheet in AdS(5) x S(5),}
Nucl. Phys. B802 (2008) 1, 
arXiv:0712.4278 [hep-th].}
and shown to exist to all orders in $\a'$.
This suggests
integrability to all orders in $\a'$.

\newsec{$d=10$ Super Yang-Mills and Superparticle.}

The aim of this section is to describe SYM by performing a first
quantization of the superparticle.

\subsec{Review of the ten-dimensional superparticle}
The action for a scalar particle in $10$ dimensions can be written as

\eqn\particle { S = \int d\tau (\dot X^m P_m + e P^2 ).}
This action has reparametrization invariance, as well as Lorentz
invariance. The indices $m$ goes from $0,{\ldots} 9$, $X^m (\tau)$
denote the particle coordinates and $P_m$ its momentum conjugate. $e$
is a Lagrange multiplier which ensures the mass-shell condition $P^2 =0$.
There is a supersymmetrical version of this action \ref\BSparticle{L.
Brink and J. Schwarz, {\it Quantum Superspace,} Phys. Lett. B100 (1981)
310.}
which can be obtained from \particle\ replacing
$\dot X^m$ by a supersymmetric combination involving coordinates for
the superspace $\t^\a$, with $\a = 1,{\ldots} 16$: 
$\dot X^m \to \Pi^m  = \dot X^m - \t \g^m \dot \t$ obtaining
\eqn\BSparticle { S = \int d\tau [\Pi^m P_m + e P^2 ].}
Since $\Pi^m$ is invariant under the supersymmetry transformation 
$\d X^m = \e \g^m \t$, $\d\t^\a = \e^\a$ with constant paramenter $\e^\a$,
then \BSparticle\ is also invariant.  By
computing the canonical momentum to $p_\a$ one obtains 
\eqn\momentum{p_\a = P_m (\g^m \t)_\a .}
Since the momentum is given in term of the coordinates, one has
constraints. By defining the Dirac constraints
\eqn\constraint{d_\a = p_\a - P_m (\g^m \t)_\a,}
one can check using the canonical Poisson bracket $\{p_\a ,\t^\b \} =
\d_\a^\b$ that the constraints satisfy the algebra
$\{d_\a , d_\b\} = -2 \g^m _{\a\b} P_m.$ In order to covariantly quantize one
should covariantly separate the first and second-class constraints, but because of the 
mass-shell condition $P^2 =0$, 
there are eight first-class and eight
second-class constraints. In order to deal with the second class constraint
one can use the light-cone gauge, therefore breaking the manifest
Lorentz invariance. However, since the aim is to have a covariant
description one should explore another possibility.

\subsec{Pure spinor superparticle}

In 1985, Siegel \ref\Siegel{W. Siegel, {\it Space-Time Supersymmetric Quantum
Mechanics,} Class. Quant. Grav.2: L95, 1985. } 
proposed the following action for the superparticle
\eqn\Siegelparticle{S = \int d\tau (\dot X^m P_m + \dot \t^\a p_\a + e
P^2 ),}
which is invariant under supersymmetry as can be easily checked by
writing it in terms of supersymmetry invariant objects
\eqn\Siegelparticle{S = \int d\tau (\Pi^m P_m + \dot \t^\a d_\a + e
P^2 ),}
where $d_\a$ is defined as above. However, this attempt didn't
succeed, roughly speaking, because it has too many degrees of freedom.
Nevertheless, it was on the right track and it led to a
pure spinor version for the superparticle
\ref\BerkovitsSuperparticle{N. Berkovits, {\it Covariant quantization
of the superparticle using pure spinors,} JHEP 0109:016, 2001,
arXiv[hep-th /0105050.]} by modifying \Siegelparticle\ to
\eqn\Berkovitsparticle{S = \int d\tau (\dot X^m P_m + \dot \t^\a p_\a
+\dot {\l^\a} \o_\a )} 
where
$\l^\a$ is a bosonic pure spinor ghost and $\o_\a$ its conjugate momentum.
Pure spinors made their first appearance in d=10 super-Yang-Mills
in \ref\bengt{
B.E.W. Nilsson, 
{\it Pure Spinors As Auxiliary 
Fields In The Ten-Dimensional Supersymmetric Yang-Mills Theory},
Class.Quant.Grav.3:L41,1986.}, and Paul Howe was the first to 
recognize that pure spinors
simplify the description of the super-Yang-Mills (and supergravity)
equations of motion
and gauge invariances \ref\howeone{P. Howe, {\it
Pure spinors lines in superspace and ten-dimensional supersymmetric theories},
Phys.Lett.B258:141-144,1991, Addendum-ibid.B259:511,1991.}\ref\howetwo
{P. Howe, {\it
Pure spinors, function superspaces 
and supergravity theories in ten dimensions and eleven dimensions,}
Phys.Lett.B273:90-94,1991.}.

An 
unconstrained spinor in ten dimensions has $16$ degrees of freedom,
but $\l$ is
constrained to satisfy the pure spinor condition $\l \g^m \l =0$.
Because of this constraint one has $11$ degrees of freedom.
Naively counting, one should have $12$ bosonic ghosts since, if one
counts the 8 fermionic second-class constraints as 4 fermionic first-class
constraints, one has a total of $12$ fermionic
first-class constraints. The fact that $\l$ only has 11 components is
because one of the 12 bosonic ghosts is cancelled by the fermionic ghost
which comes from the $P^2=0$ constraint. 
To see why a pure spinor has $11$ independent (complex)
components, note that a $U(5)$ subgroup of the (Wick-rotated) Lorentz group
leaves invariant a pure spinor up to a complex phase. So pure spinors
parameterize the space $C\times {{SO(10)}\over{U(5)}}$ which is an
eleven-dimensional complex space.
Because of the pure spinor condition, the worldsheet action is
invariant under $\d\o_\a = \L^m (\g_m \l)_\a$ which means that $\o_\a$
has 11 gauge-invariant components.

Pure spinors were first defined by Cartan \ref\Cartan{E. Cartan,
{\it Lecons sur la Theorie des Spineurs}, Hermann, Paris, 1937.}. A
product of two bosonic spinors in even dimension
$d=2D$ can be written (up to coefficients) as
\eqn\Fierz{\l^\a \l^\b = 
(\l \g^{m_1 {\ldots}
m_{D}}\l)(\g_{m_1 {\ldots} m_{D}})^{\a\b} + 
(\l \g^{m_1 {\ldots}
m_{D-4}}\l)(\g_{m_1 {\ldots} m_{D-4}})^{\a\b} + ...,}
where $(\g^{m_1 {\ldots}
m_n})_{\a\b}$ for $n = 1,{\ldots} D$ 
denotes the antisymmetrization of the $n$ indices and when $n$ is $D$
{\it mod} $4$, $(\g^{m_1 {\ldots} m_n})_{\a\b}$ is symmetric in $\a\b$.
Cartan's definition of pure spinors states that the only
nonvanishing component of this decomposition is the one involving the
$D$ form. This definition coincides with the $10$-dimensional
definition of a pure spinor given above. 

\subsec{$D=10$ Super Yang-Mills}
Although it is not known how to write an action for Super Yang-Mills in
$10$ dimensions invariant under supersymmetry transformations, it is
known how to write the equations of motion for SYM in a manifestly
covariant way. To write this equation of motion, one can use intuition
and modify the ordinary derivatives $\p_m$ and supersymmetric
derivatives $D_\a = {\p \over {\p\t^\a}} + (\g^m \t)_\a \p_m$ which
commutes with space-time supersymmetry and 
satisfy $\{D_\a , D_\b\} = 2\g^m_{\a\b}\p_m$; by 
\eqn\covariantod{\p_m \to \N_m = \p_m + A_m (X,\t),}
\eqn\covariantsd{D_\a \to \N_\a = D_\a + A_{\a} (X,\t),}
where $A_\a$ and $A_m$ are superfields. The covariant derivatives 
now satisfy $\{\N_\a , \N_\b \} =
2\g^m _{\a\b}\N_m $. The equations of motion for the superfield $A_\a$
is
\eqn\eomA{\N_\a A_{\b} + \N_\b A_{\a} + \{A_\a ,A_\b\} = 2\g^m
_{\a\b}A_m ,}
from which one gets
\eqn\eAm{A_m = {1\over 32} (\g_{m})^{\a\b} (\N_{(\a}A_{\b )} + \{A_\a
,A_\b\})}
and also 
\eqn\eomAalpha{\g_{mnpqr}^{\a\b} (\N_{(\a}A_{\b )} + \{A_\a
,A_\b\})=0.}
There is of course a gauge invariance $\d A_\a (X,\t) = \N_\a \O(X,\t)$, 
$\d A_m(X,\t) = \N_m \O(X,\t)$ and the first one can be used to gauge
fix some of the field components of $A_\a (X,\t)$, such that
\eqn\Aexpansion{A_\a (X,\t) = a_m (\g^m \t)_\a + \chi^\b (\g^m \t)_\b
(\g_m \t)_\a + \p_m a_n (\t\g^{pmn} \t)(\g_p \t)_\b +{\ldots} }
where
\eqn\eomsym{\p^m(\p_{[m}a_{n]}) =0 , \,\,\, \p^m (\g_m
\chi) =0.}

These equations of motion can be obtained as constraints by
quantizing the superparticle. If one
defines the BRST charge $Q = \l^\a D_\a$,
then it is nilpotent since $Q^2 = (\l \g^m \l)\p_m =0$ when $\l$ satisfies the
pure spinor condition $\l\g^m\l =0$.
The vertex operator will be a ghost number one operator, written in
terms of the SYM superfield as
\eqn\symvo{V = \l^\a A_\a (X,\t).}
By computing $(Q + V)^2 =0$ one encounters that $A_\a (X,\t) $ is
on-shell. The BRST operator also generates the gauge invariance for the
vertex operator $\d V = Q\O (X,\t)$ which implies $\d A_\a (X,\t) = D_\a \O
(X,\t)$.

\newsec{Pure Spinor Superstring and Tree Amplitudes}

\subsec{Worldsheet variables}

The action for the flat space superstring using the pure spinor
formalism is written as 
\eqn\flataction{S = \int d^2 z ({1\over 2}\p X^m \pb X_m + p_\a \pb \t^\a +
\o_\a \pb \l^\a + \hat p^\ah \p \hat\t^{\ah} + \hat\o_{\ah} \p \hat\l^{\ah}),}
where for the open string case one would have the boundary conditions
$\t^\a = \hat \t^\a $, $\l^\a = \hat \l^\a $.
For the Type IIA string, the $\hat\a$ spinor index has the opposite chirality
from the $\a$ spinor index, while for the Type IIB string it is of the same
chirality.
The left-moving BRST charge is given by $Q = \oint \l^\a d_\a,$
where now $d_\a$ stands for
\eqn\susyder{d_\a = p_\a + \p X^m (\g_m \t)_\a + {1\over 8}(\g^m 
\t)_\a(\t \g_m \p \t),}
and satisfies the OPE\ref\siegt{W. Siegel,
{\it Classical Superstring Mechanics,}
Nucl. Phys. B263 (1986) 93.}
\eqn\dd{d_\a (y) d_\b(z) \to {{\g^m _{\a\b}\Pi^m}\over{y-z}},}
where $\Pi^m = \p X^m - \t \g^m \p \t $. 

\subsec{Physical states}
A physical state at ghost number $1$ in the cohomology of $Q$ can be
written as 
\eqn\uvo{V = \l^\a A_\a (X,\t)}
for the massless case, while for the lowest massive case can be written as
\ref\BCmvo{N. Berkovits and O. Chandia, {\it Massive Superstring
Vertex Operator in $D=10$ Superspace}, JHEP 0208:040, 2002,
arXiv:hep-th/0204121.}
\eqn\BCmassive{V = \l^\a \Pi_m A_{\a}^m (X,\t)+\l^\a \p \t^\b A_{\a\b}
(X,\t) +\l^\a d_\b A_{\a}^\b (X,\t)} $$+ \l^\a N_{mn} 
A_{\a}^{mn} (X,\t) + \p \l^\a B_\a (X,\t) +
\l^\a J A_\a (X,\t),$$
where $N^{mn} = \half \o \g^{mn} \l$ and $J = \l^\a \o_\a$.
The central charge has a contribution of $10$ coming from the $X$'s,
$-32$ coming from $\t$, and $22$ coming from $\l$, so the total central
charge is zero. Because of the pure spinor condition, the OPE's of
$\l$ and $\o$ have to be done with care: One can do a $U(5)$
decomposition, losing manifest ten-dimensional Lorentz covariance,
but at the end, the result can be expressed in terms of the Lorentz
currents in the following covariant way

\eqn\OPENN{N^{mn}(y) N^{pq}(z) \to {{\eta^{m[p}N^{q]n} -
\eta^{n[p}N^{q]m} }\over{y-z}} - 3{{\eta^{m[p}\eta^{q]n} -
\eta^{n[p}\eta^{q]m} }\over{(y-z)^2}}.}
 
Note that
the OPE for the Lorentz currents corresponding to the
matter sector $M^{mn} = \half (p\g^{mn}\t)$ is
\eqn\OPEMM{M^{mn}(y) M^{pq}(z) \to {{\eta^{m[p}M^{q]n} -
\eta^{n[p}M^{q]m} }\over{y-z}} + 4{{\eta^{m[p}\eta^{q]n} -
\eta^{n[p}\eta^{q]m} }\over{(y-z)^2}}.}
So for the total Lorentz current $M^{mn} + N^{mn}$, the
double pole is the same as in the RNS formalism where the Lorentz
current is $\psi^m\psi^n$.

\subsec{Tree amplitudes}

The simplest case to consider is the scattering amplitude of three
open string states 
\eqn\ttscattering{\langle V_1 (z_1) V_2 (z_2) V_3 (z_3)\rangle =
\langle \l^\a A_\a ^1 (z_1) \l^\b A_\b ^2 (z_2) \l^\g A_\g ^3 (z_3)\rangle.}
After using the OPE's one is faced with the following integral
$\int d^{10}X \int d^{16}\t \int d^{11}\l$ which diverges, so one has 
to regularize it. One can use intuition from bosonic
string theory for deciding which zero
modes of $\l^\a$ and $\t^\a$ need to be present for non-vanishing
amplitudes. In bosonic string theory, the zero-mode
prescription coming from functional integration is
\eqn\bsprescription{\langle c \p c \p^2 c \rangle =1}
where $c$ is the worldsheet ghost coming from fixing the conformal gauge.
It happens that $c \p c \p^2 c$ is the vertex operator of $+3$
ghost-number for the Yang-Mills antighost \ref\Siegelfirst{W. Siegel,
{\it First Quantization and Supersymmetric Field Theories,} Stony
Brook 1991 Strings and Symmetries Proceedings (1991).}. It is
natural to use this ansatz and impose that non-vanishing correlation
functions in this formalism must also be proportional to the vertex
operator for the Yang-Mills antighost, which is $(\l \g^m \t)(\l \g^n
\t) (\l \g^p \t)(\t \g_{mnp}\t)$ \ref\Berkovitsuperparticle{N.
Berkovits, {\it Covariant Quantization of the Superparticle using Pure
Spinors,} JHEP 09 (2001), hep-th/0105050.}. So, the zero mode prescription for
tree amplitudes in the pure spinor formalism is 
\eqn\treelprescription{\langle (\l \g^m \t)(\l \g^n
\t) (\l \g^p \t)(\t \g_{mnp}\t) \rangle =1.}

Although there is a generalization of this prescription for computing
loop amplitudes which involves
picture-changing operators
\ref\multiloop{N. Berkovits, {\it Multiloop amplitudes
and vanishing theorems using the pure spinor formalism for the
superstring, } JHEP 0409:047, 2004, hep-th/0406055.}, a better
method is to
introduce a new set of ``non-minimal'' variables $\lb_\a$ and $r_\a$, with
corresponding conjugate momenta $\ob^\a $ and $s^\a$ \ref\nekn{N. Nekrasov,
{\it private communication}\semi
N. Nekrasov,
{\it Lectures on curved beta-gamma systems, pure spinors, and anomalies,}
hep-th/0511008.}.
The left-moving contribution to the
action for the non-minimal pure spinor formalism 
\ref\NMPS{N. Berkovits, {\it Pure spinor formalism as an N=2
topological string, } JHEP 0510:089, 2005. arXiv hep-th/0509120.}
is
given by 
\eqn\nmps{S = \int d^2 z ({1\over 2 }\p X^m \pb X_m + p_\a \pb \t^\a +
\o_\a \pb \l^\a +\ob^\a \pb \lb_\a + s^\a \pb r_\a).}
$\lb$ is constrained to satisfy the pure spinor condition
$\lb \g^m \lb =0$ and 
one also imposes that $\lb \g^m r =0$. Note that $\lb_\a$ and $\ob^\a$
are bosons, 
and $r_\a$ and
$s^\a$ are fermions. The BRST charge is now $Q_{nonmin} = \int dz
(\l^\a d_\a + \ob^\a r_\a)$ so that the cohomology is not modified
and
all physical states can be chosen to be independent of the new variables. 

Non-minimal pure spinor variables are useful because one can now construct a
regulator $exp(\{Q,\L\})$ which makes finite the measure of
integration. Note that the regulator is equal to $1+Q\Omega$, so
it does not affect BRST-invariant amplitudes. If one defines $\L=
-\bar\l_\a\t^\a$ so that $Q\L = -\bar\l_\a\l^\a - r_\a\t^\a$ and
inserts the regulator 
$exp(\{Q,\L\})$, the integral becomes
\eqn\measure{\int d^{10} X \int d^{16} \t \int d^{11}\l \int d^{11}\lb
\int d^{11}r f(X,\t,\l)\to}
$$
\int d^{10} X \int d^{16} \t \int d^{11}\l \int d^{11}\lb
\int d^{11}r e^{\{Q,\L\}} f(X,\t,\l) $$
$$= 
\int d^{10} X \int d^{16} \t \int d^{11}\l \int d^{11}\lb
\int d^{11}r 
 e^{- \l^\a \lb_\a
- r_\a \t^\a} f(X,\t,\l) .$$
If $\lb_\a$ is interpreted as the complex conjugate to $\l^\a$, this
choice of $\L$ regularizes the integration over $\l$.
Since $r$ does not appear 
in $f(X,\t,\l)$, one can show that \measure\ is equal to 
\eqn\rmeasure{T^{\a\b\g\d_1 ...\d_5}
\int d^{10}X \int (d^5 \t)_{\d_1 ...\d_5}
 ({\p \over {\p\l}})_{\a\b\g}^3 f(X,\t, \l)}
where the tensor
$T_{\a\b\g\d_1 {\ldots} \d_5 }$ (the inverse of
$T^{\a\b\g\d_1 {\ldots} \d_5 }$) is a Lorentz-invariant tensor defined by
\eqn\cov{(\l \g^m \t)(\l\g^n \t) (\l\g^p \t)(\t
\g_{mnp}\t)
=T_{\a\b\g\d_1 {\ldots} \d_5 }\l^\a \l^\b \l^\g \t^{\d_1}{\ldots}
\t^{\d_5}.} To obtain \rmeasure, one uses that $\bar\l\g^m r=0$
implies that 
\eqn\rcov{\int d^{11}r = T^{\a\b\g\d_1 ...\d_5}
\e_{\d_1 ... \d_{16}}
{\p \over{\p r_{\d_6} }}{\ldots} {\p
\over{\p r_{\d_{16}} } }\lb_\a \lb_\b \lb_\g.}
So \measure\ reproduces the ansatz of 
\treelprescription.

The four-point amplitude at tree level is given by considering
three unintegrated vertex operator and one integrated vertex operator
\eqn\fourptamplude{A_4 = \langle V_1 (z_1) V_2 (z_2) V_{3}(z_3) \int dz_4
U(z_4)\rangle .}
To find the form of the integrated vertex operator $U$,
start with the superparticle action 
\eqn\sparticle{\int d \tau (\dot X^m P_m + \dot \t^\a p_\a + \o_\a
\dot \l^\a),}
and consider a super Yang-Mills background
\eqn\symbackground{\int d\tau (\dot X^m P_m + \dot \t^\a p_\a + \o_\a
\dot \l^\a + e(A_m \dot X^m + A_\a \dot \t^\a + {\ldots} ))}
where $...$ is determined from BRST invariance.
In RNS, the integrated operator
is  $\int d\tau (A_m \p X^m + \psi^m \psi^n \p_n A_m)$ where the
last term is determined by worldsheet superconformal invariance. 
In the pure spinor formalism, the
integrated vertex operator is determined by BRST invariance and is given by 
\eqn\integrated{ U = A_m \Pi^m + A_\a \p \t^\a + W^\a d_\a +
F^{mn}N_{mn},}
where $W^\a$ and $F_{mn}$ are superfield strengths. The lowest
component of $W^\a$ is the gaugino $\chi^\a$ and
the lowest component of $F_{mn}$ is the fieldstrength 
$\p_{[m}a_{n]}$. One can check that
$Q U = \p (\l^\a A_\a)$ so $\int dz U$ is BRST invariant.

The N-point tree level amplitude
\eqn\Npttree{\langle V^1 (z_1) V^2 (z_2) V^3 (z_3) \int U_4 {\ldots}
\int U_N \rangle}
can be computed by first integrating out the non-zero modes by
evaluating the OPE's. To integrate the zero modes, use

\eqn\intzeromodes{\langle f (X,\t,\l)\rangle = T\int d^{10} X ({\p\over
{\p\l}})^3  ({\p\over
{\p\t}})^5 f} 
where $T$ is the tensor of \cov. From the three point 
tree level amplitude 
$\langle \l A \l A \l A \rangle$ 
one gets the usual cubic term in the SYM amplitude $\int d^{10}X (a a
\p a + \chi a \chi)$.

\newsec{Loop Amplitudes}

\subsec{$b$ ghost}

In the pure spinor formalism there is no analog of the $c$ ghost, but
there is an analog of the $b$ ghost which is necessary for the
computation of string loop amplitudes.
For example, the closed string one loop amplitude requires
a $b$ and $\bar b$ ghost integrated over the Beltrami
differential of the torus as
\eqn\Beltramint{\int d^2 \tau \langle V_1 \int b \int \bar b \int U_2
{\ldots} \int U_N \rangle}
where in the case of the closed string, $V = \l^\a \hat \l^{\bh}
A_{\a\bh}
(X,\t ,\hat \t)$. Note that 
at the linearized level, BRST invariance of this vertex operator implies
that $A_{\a\bh}$ satisfies the supergravity
equations of motion.

It will be shown that a composite operator for the $b$ ghost can be
written in terms of the other worldsheet fields in
such a way that $\{Q,b\} = T$. To construct this operator, note that after
adding the non-minimal variables of the previous section, the energy 
momentum tensor is given by 
\eqn\EM{ T_{nonmin} = {1\over 2}\Pi^m \Pi^m + d_\a \p \t^\a + 
s^\a \p r_\a +  T_\l + T_{\lb} }
where $T_\l$ and $T_{\lb}$ are the stress tensors for $\l^\a$ and $\lb_\a$.
If one would start with $b^\a = {1\over 2}\Pi^m (\g^m d)^\a $, then $Q b^\a
= {1\over 2}\Pi^2 \l^\a$ up to terms involving $\p\t^\a$. 
So, naively, one should ``divide'' $b^\a$ by $\l^\a$. 
With the help of the non-minimal variables, this is possible by
defining
\eqn\bdef{b =
{{1\over 2 }\lb_\a (\Pi^m \g_m d)^\a \over{\l^\b \lb_\b}} + ... }
where $...$ is determined by $\{Q_{nonmin},b\}=T_{nonmin}$ where
$Q_{nonmin} =\int dz (\l^\a d_\a + \bar\o^\a r_\a)$.
One finds that the complete expression for the $b$ ghost is 
\eqn\nonminb{b =  s^\a\p\lb_\a  + {{\lb_\a (2
\Pi^m (\g_m d)^\a-  N_{mn}(\g^{mn}\p\t)^\a
- J_\l \p\t^\a -{1\over 4} \p^2\t^\a)}\over{4(\lb\l)}} } 
$$+ {{(\lb\g^{mnp} r)(d\g_{mnp} d +24 N_{mn}\Pi_p)}\over{192(\lb\l)^2}} 
- {{(r\g_{mnp} r)(\lb\g^m d)N^{np}}\over{16(\lb\l)^3}} +
{{(r\g_{mnp} r)(\lb\g^{pqr} r) N^{mn} N_{qr}}\over{128(\lb\l)^4}}, $$
which satisfies $\{Q_{nonmin} , b\} = T_{nonmin}$. From 
now on, the $nonmin$ subscript will be dropped out.

The fact that the $b$ ghost has poles when $\l\lb \to 0$ means
there are subtleties in defining the Hilbert space of allowable
states in the pure spinor formalism. If one allowed states with
arbitrary powers of poles, the cohomology would become trivial.
This is easy to verify since the operator 
\eqn\con{S = {\t \lb \over{\l \lb + r \t}},}
satisfies $Q S =1$. Then $QV = 0$ implies $Q (SV) = V$, so
the existence of $S$ in the Hilbert space would trivialize the BRST cohomology.
Expanding $S$, one finds a pole of 11th order when $(\l\lb)=0$. So
if one allowed operators with this pole behavior in $\l\lb$, the
cohomology would become trivial. One therefore forbids states
in the Hilbert space which diverge faster than $(\l\lb)^{-10}$
when $\l\to 0$. This allows the above operator for the $b$ ghost 
but forbids the $S$ operator.

\subsec{Loop amplitude computations}

For $g$-loop amplitudes, one needs to insert $3g-3$ $b$ ghosts.
So for $g\geq 3$, 
the number of poles in the $b$ ghost could add up to more than 11.
This would make the functional integral $\int d^{11}\l \int d^{11}
\bar\l$ diverge near $\l\lb=0$.
This difficulty is overcome with an appropriate
definition of a regulator \ref\BN{N. Berkovits and N. Nekrasov, {\it 
Multiloop superstring amplitudes from non-minimal pure spinor
formalism,} JHEP 0612:029, 2006, arXiv:hep-th/0609012.}
which smooths out the poles of the different $b$ ghosts so that
the total divergence is slower than $(\l\lb)^{-11}$.
However, the form of this regulator is complicated and its
explicit contribution has only been worked out in simple
cases 
\ref\AB{Y. Aisaka and N. Berkovits, {\it Pure Spinor Vertex
Operators in Siegel Gauge and Loop Amplitude Regularization, } JHEP
0907:062, 2009, [arXiv:0903.3443].}. 
Nevertheless, there are several multiloop amplitudes
one can compute which do not require this complicated regulator.

In the non-minimal pure spinor formalism, the integration measure at
$g$ loops is  
\eqn\amplitude{{\it A }= \int d^{10} X \int d^{16}\t \int d^{11}\l \int
d^{11}\lb \int d^{16g} p \int d^{11g}\o \int d^{11g}\ob \int d^{11g}s \int
d^{11}r }
where the conformal weight one worldsheet fields contribute with $g$
zero modes. One can separate out the non-zero modes
and use the free field OPE's to integrate them out,
leaving an integration over bosonic and fermionic zero modes.
To account for the bosonic and fermionic zero modes,
the zero mode regulator used for tree-level amplitudes 
must be
modified to $\L = -\lb_\a \t^\a - \sum_{I=1}^g\o_{I\a} s_I^\a$ which implies 
$Q\L = \lb_\a \l^\a - r_\a \t^\a - 
\sum_{I=1}^g (\ob_I^\a \o_{I\a} - s_I^\a d_{I\a})$ \NMPS.

As an example, one can compute the four-point massless one-loop and
two-loop amplitudes. Using \Beltramint, the one-loop four-point open
superstring amplitude is given by 
\eqn\fourpt{{\it A} = \int d^2 \tau \int d^{10} X \int d^{16}\t \int d^{11}\l \int
d^{11}\lb \int d^{16} p \int d^{11}\o \int d^{11}\ob \int d^{11}s \int
d^{11}r \int b  }$$ (\l A) (\int \p \t^\a A_\a + \Pi^m A_m  + d_\a W^\a +
N^{mn}F_{mn})^3 e^{\{Q,\L \}}.$$
To get a non-vanishing amplitude,
one needs to absorb $16d_\a$ zero modes from $\int
d^{16}p$. One can get $3$ from the term $d_\a W^\a$. The maximum number
of $d_\a$ zero modes one
can get from the regulator is $11$, so the remaining two must come from the
third term in the $b$ ghost. This third term of $b$ has one $r_\a$, so
the remaining $10$ $r$'s must come from the regulator. Note that
$\o$ $\ob$ and $\l\bar\l$ have gaussian integrals, which are easy to
compute. So after integrating over the zero modes of $p_\a$, $r_\a$
and $s^\a$, one finds a term proportional to 
\eqn\answer{\int d^{16}\t~ \t^{10}AWWW }
where the factor of $\t^{10}$ comes from the regulator, and indices
on the superfields in \answer\ are contracted in a Lorentz-invariant
manner.
The computation of the Lorentz index contractions for
the gluon contribution was done
in \ref\Mone{C. R. Mafra, {\it
Four-point one-loop amplitude computation in the pure spinor formalism,}
JHEP 0601 (2006) 075, 
hep-th/0512052.}, giving as a result $t_8 f^4$ where $t_8$
is a Lorentz-invariant tensor which contracts the 8 indices of $f^4$.
For closed strings the analogous result was $t_8 t_8 R^4$. Using the
non-minimal pure spinor formalism, the gauge anomaly one loop
computation was also performed in \BM, and five point one 
loop computations  were performed in
\MafraStahn.

For four point two-loops, the closed string amplitude is given by
\eqn\amplitudetwol{{\it A} =  \int (d^2\tau)^3  \langle
(\int b)^3 (\int
\bar b)^3 \int U_1{\ldots} \int U_4 e^{\{Q,\L\}}\rangle .}
Because of the two non-trivial cycles, 
\eqn\regulator{\L = - \lb_\a \t^\a - \displaystyle\sum_{I=1}^2  \o_\a
^{(I)}s^{\a (I)},}
and
\eqn\Qregulator{\{Q,\L \} = - \lb_\a \l^\a - r_\a \t^\a -
\displaystyle\sum_{I=1}^2 (\ob^{\a (I)}\o_\a
^{(I)} - s^{\a (I)}d_\a ^{(I)}).}
One now needs to absorb $32 d_\a$ zero modes. The regulator contributes 
$22$, each vertex operator contributes $1$ and, because there are three
$b$ fields, the third term in \nonminb\ gives the remaining $6$ and
also absorbs 3 $r_\a$ zero modes. The regulator 
absorbs the 22 $s^\a$ zero modes and also absorbs the remaining 8 $r_\a$
zero modes and contributes 8 $\t^\a$ zero modes. So the resulting amplitude
is of the form 
\eqn\derf{|~\int d^{16} \t~ \t^8 WWWW~|^2 .}
The Lorentz index contractions for the graviton contribution was shown in
\twoloop\ 
to give $t_8 t_8 \p^4 R^4$, and confirmed the Type IIB S-duality
prediction \ref\greend{M.B. Green and M. Gutperle, {\it
Effects of D instantons,} Nucl. Phys. B498 (1997) 195, 
hep-th/9701093\semi
M.B. Green and P. Vanhove, {\it D-instantons, strings and M-theory,}
Phys. Lett. B408 
(1997) 122, hep-th/9704145.} that $\p^4 R^4$ is the term of lowest
order in derivatives at two loops. 

\subsec{Non-renormalization theorems}

Now one can ask what is the term of lowest order in derivatives 
at higher loops. At $g$ loops, the naive expression for the term
of lowest order in derivatives which saturates the fermionic zero modes
is 
\eqn\ampmultiloop{{\it A} =  \int d^{16}\t 
\int d^{11} r  \int d^{16g} p \int d^{11g}s ~ (r\t)^{12-2g}(ds)^{11g}
(rdd)^{2g-1} (\Pi d)^{g-2} ~d^4,} 
where $(r\t)^{12-2g} (ds)^{11g}$ comes from the regulator, 
$(rdd)^{2g-1} (\Pi d)^{g-2} $
comes from the $3g-3$ $b$ ghosts, and $d^4$ comes from the four
vertex operators. This naive formula predicts that the term of
lowest order in derivatives at $g$ loops is
$|~\int d^{16}\t (\t)^{12-2g} WWWW~|^2$, which corresponds to
$\p^{2g} R^4$.
However, this formula clearly breaks down at $g>6$ because of the 
$(r\t)^{12-2g}$ term in \ampmultiloop. 

The source of this breakdown is the divergence when $\l\lb\to 0$.
For $g<6$, one can argue that these divergences are not present since
the terms in the $b$ ghost which contribute do not diverge faster than
$(\l\lb)^{-10}$. This is related to the fact that $\p^{2g} R^4$ is a superspace
$F$-term when $g<6$. However, when $g\geq 6$, the poles from the $b$
ghost diverge faster than $(\l\lb)^{-10}$ which means one has to use
the complicated regulator of \BN. This is related to the fact that
$\p^{2g} R^4$ can be written as a superspace $D$-term when $g\geq 6$.
In the presence of the complicated regulator, the zero mode counting
of \ampmultiloop\ is modified. Although a detailed analysis of the
zero mode counting has not yet been done in the presence of this
complicated
regulator, it naively appears that the $\p^{12} R^4$ term can appear at 
all loops for $g\geq 6$ \nrtheorems. 
If this naive counting is correct, it would
imply (by dimensional arguments) that the first divergence of $N=8$
d=4 supergravity appears at 9 loops \GRV.

\newsec{Curved Backgrounds}

\subsec{$\a'$ corrections to supergravity}

The action in a curved
background can be obtained by considering the flat background with
vertex operators, and then covariantizing.
Use the variables $Z^M = (X^m , \t^\mu)$ for the open string. In this
notation, $\p \t^\a A_\a + \Pi^m A_m$ combines to $\p Z^M A_M$. For
the closed superstring, use the coordinates $(X^m, \t^\mu, \bar\t
^{\hat\mu})$. 
One gets the action\ref\hb{N. Berkovits and P. Howe,
{\it Ten-dimensional supergravity constraints 
from the pure spinor formalism for the superstring},
Nucl.Phys.B635:75-105,2002,
hep-th/0112160.}
\eqn\generalbackground{S = \int dz d\bar z ( {1\over 2}(G_{MN}+B_{MN})\p Z^M \pb
Z^N + E_{M}^\a d_\a \pb Z^M + E_{M}^{\ah} \bar d_{\ah} \p Z^M +
F^{\a\bh}d_\a\bar d_{\bh}} $$+
\O_M^{ab} \p Z^M \bar N_{ab} 
+
\bar\O_M^{ab} \bar\p Z^M N_{ab} +
C^{\a ab}d_\a \bar N_{ab}+\bar C^{\ah ab}\bar
d_{\ah} N_{ab} + R^{abcd}N_{ab}\bar N_{cd}+\o_\a \pb \l^\a +
\ob_{\ah}\p \lb^{\ah} ).$$
The index notation is $A = (a,\a,\ah)$ and $E_M^A (Z)$ is the
supervielbein. Note that the superspace metric $G_{MN} = E_M ^a E_N ^b
\eta_{ab}$ does not involve the supervielbein with indices
$(\a,\ah)$. So all the components of $E_M ^A (Z)$ appear in
the action, while in the Green-Schwarz action $E_M ^\a (Z)$ and
$E_M^\ah(Z)$ do not
appear. In \generalbackground, the lowest component of $F^{\a\bh}$ is the Ramond-Ramond field
strength. 
Note
that $d_\a$ is treated as an independent variable in this action instead
of $p_\a$.

To compute $\a'$ corrections to
the supergravity equations of motion using this action,
one should compute whether the action is BRST
invariant, or equivalently, if the BRST charge $Q$ is nilpotent
and conserved.
It was shown in \hb\ that nilpotence of $Q$ and
$\pb (\l^\a d_\a) =0$ at the classical
level implies the
supergravity equations of motion to lowest
order in $\a'$. These equations of motion imply $\kappa$-symmetry
in the Green-Schwarz formalism. Hovever, because $E_M ^\a$ does not
appear explicitly in the action in the GS formalism, it is 
not true that $\kappa$-symmetry implies the supergravity equations of motion. 

At higher loop order, one needs to introduce the dilaton coupling $\a' \int d^2
z \Phi (Z) r$ and compute loop corrections
to the OPE of $Q$ with $Q$ and the OPE of the stress
tensor with $Q$.
The one-loop
Yang-Mills Chern-Simons corrections have been computed in this manner
\ref\YMCS{O.
Bedoya, {\it Yang-Mills Chern-Simons Corrections from the Pure Spinor
Superstring, } JHEP 0809:078,2008, arXiv:0807.3981 [hep-th]}.

\subsec{$AdS_5\times S^5$ background}

If $F^{\a\bh}$ is an invertible matrix as in
the $AdS_5 \times S^5$ background,
one can solve the auxiliary
equations of motion of $d_\a$ and write $d_\a$ in terms
of $Z^M$.
Because of $PSU (2,2|4)$ isometry in this background,
it is natural to define $E_M ^A$ as in 
\ref\MetsaevTseytlin{R. Metsaev and
A. Tseytlin, {\it Type IIB Superstring Action in $Ads_5 \times S^5$
Background,} Nucl. Phys. B533 (1988) 109, hep-th/980502.}
in terms of a coset $g(z) \in
{{PSU(2,2|4)}\over{SO(4,1)\times SO(5)}}  \simeq {{SO(4,2)\times
SO(6)}\over{SO(4,1)\times SO(5)}} + 32$ fermions. The left-invariant
currents are
defined by $J = (g^{-1} \p g)$ and  $\Jb = (g^{-1} \pb g)$ where the global
$PSU(2,2|4)$ isometries act on the left as $g \to \Sigma g$. The action
will be defined to be
invariant under local transformations by the right $g \to g \O(z)$
where $\O(z)$ takes values in $SO(4,1)\times SO(5)$.

The currents can be decomposed into the ten vector elements
$J^a$ and $J^{a'}$ (where $a= 0,{\ldots} 4 , a'=5...9$),
the 32 fermionic elements $J^\a$ and $J^{\ah}$ (where $\a,\ah =
1{\ldots} 16$), and the 20 bosonic elements
$J^{[ab]}$ and $J^{[a'b']}$, where $[ab] \in SO(4,1)$ and $[a'b']\in SO(5)$. 
These currents can also be written in terms of the
vielbeins as $J^A = E_M ^A \p Z^M$, where $E_M ^{[ab]}$ is defined to be the
spin connection $\Omega_M ^{[ab]}$.
After using the equations of motion to solve for $d_\a$ and $\bar d_{\ah}$,
the BRST charge can be written as 
\eqn\brstchargeads{Q = \int dz \l^\a J^{\ah}\eta_{\a\ah} + \int d\bar z
\lb^{\ah}\Jb^\a \eta_{\a\ah},}
where $\eta_{\a\ah} = (\g^{01234})_{\a\ah}$ is in the direction of the
RR field strength.

The pure spinor action in the $AdS_5\times S^5$ background can be written as 
\eqn\Adsaction{S = R^2 \int d^2 z ({1\over 2 } J^{a} \Jb _{a} +{1\over 2}
J^{a'}\bar J_{a'} +
 \d_{\a\bh} (J^\a \Jb^{\bh} - 3 J^{\bh} \Jb ^\a) +
\o_\a \Nb \l^\a +\ob_{\ah} \N \lb^{\ah}}
$$+ (\o \g^{ab}\l)(\ob \g_{ab}\lb) -
(\o \g^{a'b'}\l)(\ob \g_{a'b'}\lb) ) ,$$
where the last line appears because the space-time curvature of $AdS_5
\times S^5$ is non-vanishing. 

To show that this action has BRST symmetry, note that the BRST
charges act on the group elements as $Q g = g (\l^\a T_\a + \lb^{\ah}
T_{\ah})$ where $T_\a$ and $T_\ah$ are the 32 fermionic generators of
$PSU(2,2|4)$. From this, 
it is trivial to work out how $Q$ acts on $J$. Note that
$Q^2$ acting on $g$ will be zero because of the pure spinor condition satisfied
by $\l^\a$ and $\lb^{\ah}$. 

What can be done with this model, which looks rather simple?
One interesting question is if 
there is a version of this action which is BRST invariant to all order
in $\a'$? This can be answered in the affirmative by using cohomology arguments
\Quantumconsistency.

Since the BRST operator is nilpotent, one can ask about its cohomology.
At the lowest order in $\a'$, define the classical
action of \Adsaction\ to be $S_0$. This action
is BRST invariant since $Q  S_0 =0$. In other words, the BRST
transformation of the corresponding Lagrangian $L_0$ is a total
derivative $Q L_0 = d \Lambda_0$. After computing the quantum part of
the effective action $S_1$, one can ask
if the sum of the classical and quantum action is
still BRST invariant? In other words, is $Q(S_0 + \a' S_1) =0$, or 
equivalently, is
$Q(L_0 + \a'
L_1) = d\Lambda$?. Now, the BRST variation of the quantum 
effective action should be a local operator, since quantum anomalies come from a
short-distance regulator. Therefore,
$Q L_1 = \O_1$ where $\O_1$ is some local quantity.
Furthermore, $\O_1$ is BRST-closed since $Q^2 L_1=0$. 
One can therefore ask if $\O_1$ is BRST-exact, that is, does
$\O_1 = Q \Sigma$ for some local
$\Sigma$?. The answer happens to be yes, since the cohomology is
trivial at ghost number 1 for operators of non-zero conformal weight. 
This trivial cohomology is easily confirmed by constructing the most
general operator of ghost
number 1 which is local and which is invariant under $PSU(2,2|4)$.
Since $Q (L_0 +\a' L_1) = d
\Lambda + \a'\O_1  = d\Lambda + \a' Q \Sigma$, one can always add a
local $PSU(2,2|4)$ invariant counter-term $-\a'\Sigma$ to the Lagrangian
such that $Q(L_0 + \a' L_1 - \a'\Sigma) = d\Lambda$.
So after including the counter-term, the action 
$S_0+\a' S_1 - \a' \int d^2 z \Sigma$
is BRST-invariant. This type of argument for quantum BRST invariance
can be repeated to all perturbative orders
in $\a'$.
However, in principle there could be BRST anomalies which are 
non-perturbative in $\a'$.

The existence of non-local conserved currents is important for
integrability. The local $PSU(2,2|4)$ conserved
charges are the 
N\"{o}ether charges for the global symmetry algebra, 
\eqn\current{q^A  = \int d \sigma j^A,}
where $A$ is a $PSU(2,2|4)$ Lie algebra index.

Suppose the theory is on the plane and define the non-local charge
\eqn\nlcc{k_{(1)}^C = f_{AB}{}^C \int_{-\infty} ^\infty d\sigma j^A
(\sigma)\int_{-\infty}
^\sigma d\sigma' j^B (\sigma') - \int_{-\infty} ^\infty d\sigma l^C
(\sigma) }
for some $l^C$ where $f_{AB}^C$ are the $psu(2,2|4)$ structure constants.
Note that $Q j^A = \p_\sigma  h^A$ for some $h^A$
because $Q \int_{-\infty} ^\infty
d\sigma j^A (\sigma) =0$. Therefore,

\eqn\Qnlcc{Q k^C _{(1)}
= 2 f_{AB}{}^C \int_{-\infty} ^\infty d\sigma j^A (\sigma)h^B (\sigma) - 
\int_{-\infty}^{\infty} d\sigma Q l^C.}
So if $l^C (\sigma)$ is defined such that $Ql^C (\sigma) = 2f_{AB}^C j^A
h^B (\sigma)$, then $k_{(1)}^C$ will be BRST invariant. Using cohomology
arguments similar to those above, one can prove that there always
exists such an $l^C(\sigma)$. Therefore, one
can contruct non-local BRST conserved charges. Furthermore,
by repeatedly commuting
$k_{(1)}^C$ with each other, 
one can obtain an infinite set of conserved charges and
prove that the construction is valid at the quantum level to all
orders in perturbation theory \Quantumconsistency.
Classical non-local conserved currents have been constructed in  
\BRP\flatcurrents\mik\ and
it would be
interesting to compute the algebra of these currents.

\newsec{Open Problems}

1) Geometrical principles: At the moment, 
there is no covariant derivation of the pure spinor BRST operator 
from gauge fixing a more symmetrical formalism. Although there are various
procedures 
\ref\tonin{
M. Matone, L. Mazzucato, I. Oda, D. Sorokin and M. Tonin, {\it 
The Superembedding Origin of the Berkovits Pure Spinor Covariant 
Quantization of Superstrings,}
Nucl.Phys.B639:182-202,2002, hep-th/0206104.}
\ref\kaz{
Y. Aisaka and Y. Kazama, {\it
Origin of pure spinor superstring,}
JHEP 0505:046,2005, 
hep-th/0502208.}
\ref\expl{N. Berkovits, {\it 
Explaining the Pure Spinor Formalism for the Superstring,}
JHEP 0801:065,2008, 
arXiv:0712.0324 [hep-th].}
for getting the pure spinor BRST operator from gauge-fixing, none of
these procedures
are Lorentz covariant at all stages in the gauge-fixing.
Such a covariant derivation of the BRST operator would probably also
provide a ``geometric'' explanation of the complicated form of the $b$ ghost
\ref\ods{I. Oda and M. Tonin, {\it
Y-formalism and $b$ ghost in the 
Non-minimal Pure Spinor Formalism of Superstrings,}
Nucl.Phys.B779:63-100,2007, arXiv:0704.1219.}.
An interesting open question is to compute the cohomology of the $b$ ghost.

2) Superstring field theory: $QV + V * V=0$ where $*$ is the star product in
Witten's action gives the correct open superstring field 
theory equations of motion.
In bosonic string theory, this comes from the action $S=\langle \half
VQV + {1\over 3} V * V * V
\rangle$ \ref\wcs{E. Witten, {\it 
Noncommutative Geometry and String Field Theory,}
Nucl. Phys. B268 (1986) 253.}. 
Although $\langle \,\, \rangle$ can be defined in the
non-minimal formalism using functional integration, the expression
\eqn\defff{\langle f \rangle = 
\int d^{10} X \int d^{16} \t \int d^{11}\l \int d^{11}\lb
\int d^{11}r e^{\{Q,\L\}}f(X,\l,\t)}
only makes sense if $f$
does not have poles which diverge faster than
$(\l\lb)^{-10}$. One can insert a
regulator, but $f$ is not BRST closed since string fields are off-shell.
So the action will depend on where one puts the regulator. Furthermore,
the regulator breaks manifest spacetime supersymmetry. So although the
equations of motion are manifestly spacetime supersymmetric, the action
is not. Furthermore, to compute the four-point tree
amplitude in string field theory, one needs to introduce the $b$ ghost which
contains poles when $\l\to 0$. It is unclear how to define the off-shell
Hilbert space of allowed string fields in such a way that the product of these
string fields never contain poles which diverge faster than
$({\l\lb})^{-10}$. 

3) Multiloop amplitudes: Computations beyond two-loops require a complicated
regulator since the $b$ ghosts contribute poles which diverge faster than
$({\l\lb})^{-11}$. Up to now, no non-vanishing computations have
been performed beyond two loops. A related question is the computation
of $N$-point
tree amplitudes in a gauge which involves  more than 6 $b$ ghosts. 
These tree amplitude computations will also require the complicated regulator.

4) Unitarity: There is not yet a proof that BRST invariance of the
scattering amplitudes implies that the amplitudes are unitary.
This could be done either by proving equivalence to the RNS computation
or by proving equivalence to the light-cone GS computation.

5) Compactification: Compactifications of the pure spinor formalism on
a Calabi-Yau manifold have recently been considered in \ref\brenno{
O. Chandia, W. Linch III and B.C. Vallilo, 
{\it Compactification of the Heterotic Pure Spinor Superstring I,}
arXiv:0907.2247 [hep-th].}.
One expects
that the resulting formalism should be equivalent with the hybrid
formalism, however, this has not yet been proven. A related question
is if one can construct lower-dimensional versions of the pure spinor formalism
\ref\wyll{P.A. Grassi and N. Wyllard, {\it
Lower-dimensional pure-spinor superstrings,}
JHEP 0512 (2005) 007, hep-th/0509140.}
\ref\ido{I. Adam, P.A. Grassi, L. Mazzucato, Y. Oz and S. Yankielowicz,  
{\it Non-Critical Pure Spinor Superstrings,}
JHEP 0703 (2007) 091,
hep-th/0605118.}.

6) M-theory: There is a d=11 version of the pure spinor formalism
for the superparticle which describes linearized d=11 supergravity \ref
\mberk{N. Berkovits, {\it
Towards Covariant Quantization of the Supermembrane},
JHEP 0209 (2002) 051, hep-th/0201151.}.
The $d=11$ pure spinor is  $\l^A$, $A
=1,{\ldots} 32$ such that $\l \g^M \l =0$ for $M = 0,{\ldots} ,10$. Just
as
$Q = \l^\a d_\a$ at ghost number 1 gives SYM in 10 dimensions, $Q
= \l^A D_A$ at ghost number 3 gives linearized $d=11$ sugra. The
vertex operator at ghost number 3 is $\l^A
\l^B \l^C B_{ABC}$ where $B_{ABC}$ is the spinor component of the 3-form.
This
works nicely for the superparticle, but not has yet been generalized for the
supermembrane. The main complication is that 
the constraint $\l\g^M\l=0$ does not commute with the Hamiltonian
and
generates secondary constraints.

\vskip15pt
{\bf Acknowledgements:} OB and NB would like to thank the organizers of
``New Perspectives in String Theory'' for a very enjoyable
school and partial financial support. OB would also like to thank
the Aspen Center of Physics for hospitality during the ``Unity of
String Theory'' workshop and
FAPESP grant 09/08893-9 and CNPq grant
150172/2008-7 for partial financial support.
NB would like to thank FAPESP grant 09/50639-2 and
CNPq grant 300256/94-9 for partial financial support.

\listrefs

\end